\documentclass{PoS}

\usepackage[utf8]{inputenc}
\usepackage{amsmath}
\usepackage{amssymb}
\usepackage[super]{nth}
\usepackage{graphicx}
\usepackage{subcaption}

\newcommand{\secdec}{\texttt{\textsc{SecDec}}}
\newcommand{\secdecthree}{\texttt{\textsc{SecDec\,3}}}
\newcommand{\pysecdec}{\texttt{py{\textsc{SecDec}}}}
\newcommand{\python}{{\texttt{python}}}
\newcommand{\form}{{\texttt{FORM}}}
\newcommand{\cpp}{{\texttt{C++}}}
\newcommand{\dreadnaut}{{\texttt{dreadnaut}}}
\newcommand{\CUBA}{\texttt{\textsc{CUBA}}}

\DeclareMathOperator{\U}{\mathcal{U}}
\DeclareMathOperator{\F}{\mathcal{F}}

     \title{SecDec: a toolbox for the numerical evaluation of multi-scale integrals}
\ShortTitle{SecDec: a toolbox for the numerical evaluation of multi-scale integrals}

\author{\speaker{Stephan Jahn}\\
       Max-Planck-Institut für Physik, München, GERMANY\\
       E-mail: \email{sjahn@mpp.mpg.de}}

\abstract{
We present a new version of \secdec{}, a program for the numerical computation of
parametric integrals in the context of dimensional regularization.
By its modular structure, the \python{} rewrite \pysecdec{} is much more
customizable than earlier versions of \secdec{}. The numerical integration
is accelerated using code optimization available in \form{}. With the new \cpp{}
interface, \pysecdec{} can provide numerical solutions of analytically unknown
integrals in user-defined code.
}

\FullConference{13th International Symposium on Radiative Corrections (Applications of Quantum Field Theory to Phenomenology)\\
         25-29 September, 2017 \\
         St. Gilgen, Austria}

\begin{document}

%%%%%%%%%%%%%%%%%%%%%%
\section{Introduction}
%%%%%%%%%%%%%%%%%%%%%%

Future runs of the LHC and the subsequent high-luminosity upgrade will allow for New Physics searches mostly via precision tests of the standard model
throughout the next few decades. Higher order calculations are crucial for the upcoming data-taking to be meaningful.
A well established procedure for multi-loop calculations is: (i) draw all contributing Feynman diagrams, (ii) insert the Feynman rules to produce
algebraic expressions, (iii) identify the loop integrals to be computed, (iv) reduce these integrals to a set of master integrals, and (v) solve the
master integrals.

Analytical solutions for the master integrals are desirable and recent developments made a huge class of integrals analytically
tractable. However, phenomenologically relevant topologies, for example some nonplanar double-boxes with massive propagators, remain inaccessible with the
currently available analytical methods.

Sector decomposition~\cite{Binoth:2000ps,Heinrich:2008si} provides a numerical alternative when analytic solutions are out of reach with current techniques.
A prominent example for the successful application of this method is the recent calculations of the full top-mass dependence in Higgs boson pair
production~\cite{Borowka:2016ehy,Borowka:2016ypz}. Further studies~\cite{vonManteuffel:2017myy} indicate that picking a
finite basis~\cite{vonManteuffel:2015gxa,Panzer:2014gra,vonManteuffel:2014qoa} makes the numerical approach competitive with analytical solutions.

In this article, we present \pysecdec{}~\cite{Borowka:2017idc}, the successor of the program \secdec{}~\cite{Carter:2010hi,Borowka:2012yc,Borowka:2015mxa}.
Other public implementations of sector decomposition are \texttt{sector\_decomposition}~\cite{Bogner:2007cr} supplemented with \texttt{CSectors}~\cite{Gluza:2010rn}
and \texttt{FIESTA}~\cite{Smirnov:2008py,Smirnov:2009pb,Smirnov:2013eza,Smirnov:2015mct}.

The structure of these proceedings is as follows: In Section~\ref{sec:newfeatures} we list the key new features of \pysecdec{} compared to \secdecthree{}.
In Section~\ref{sec:usage} we illustrate the workflow from defining an integral to a sufficiently precise numerical result, including instructions
how to tune numerical integration on failure. Finally, we close with concluding remarks in Section~\ref{sec:conclusion}.

%%%%%%%%%%%%%%%%%%%%%%
\section{New features}
%%%%%%%%%%%%%%%%%%%%%%
\label{sec:newfeatures}

The major rewrite in \python{}, \form{}, and \cpp{} is completely independent of expensive commercial programs.
Rather than a text file, \pysecdec{} creates a \cpp{} library that can be used
to dynamically evaluate loop integrals occurring in amplitude calculations.

The time needed for the numerical integration
is significantly reduced by optimizing the integrand functions with \form{}~\cite{Kuipers:2013pba,Ruijl:2017dtg}.

Another significant speedup in the numerics could be achieved by identifying sectors that are equal up to permutations of the
integration variables (Feynman parameters). Instead of integrating all equivalent sectors, only one of them is numerically
integrated while the others are accounted for by multiplying the number of equivalent sectors. We implement two independent algorithms
to identify equivalent sectors: The algorithm described in~\cite{Pak:2011xt} and an algorithm based on finding graph isomorphisms with
\dreadnaut{}~\cite{MCKAY201494}, see~\cite{Jones2017:ACAT} for details.

The sign of the imaginary part of the \nth{2} Symanzik polynomial $\F$ is now checked at every point in Feynman parameter space to make sure
that the Feynman prescription is always fulfilled. Furthermore, we also check that the real part of the \nth{1} Symanzik polynomial $\U$ stays
positive on the entire contour, which is important if $\U$ has a negative exponent and in higher orders of the $\epsilon$-expansion where
$\log\U$ appears.

\begin{figure}[htb]
\vspace{5mm}
\begin{center}
\includegraphics[width=0.4\textwidth]{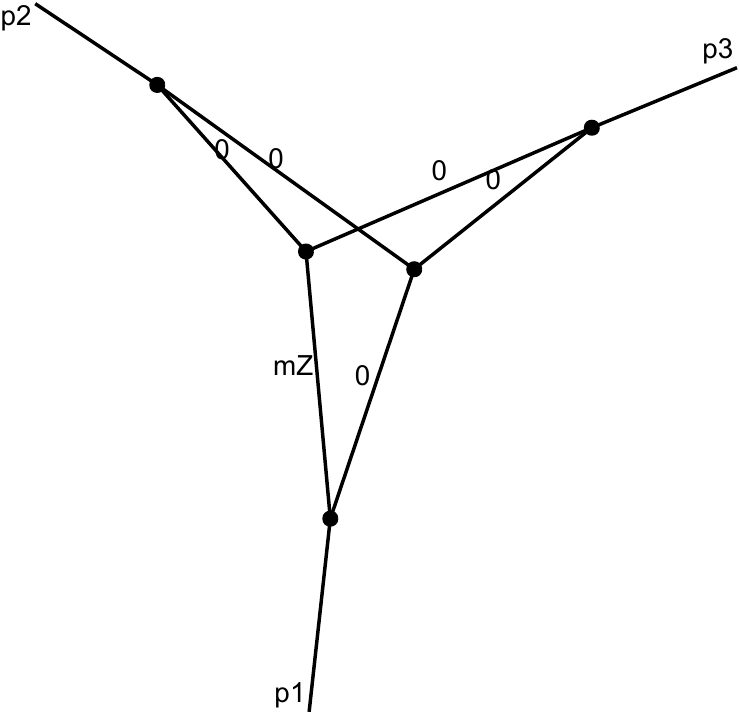}
\end{center}
\caption{Three-point function with one massive propagator ($m_Z$) and $s=p_3^2=m_Z^2$.}
\label{fig:Zbb}
\end{figure}

An improved treatment of
singularities originating from $x_i=1$ allows for the computation of integrals \secdecthree{} could not handle, e.g. the three-point
function depicted in Figure~\ref{fig:Zbb}, which occurs in single-Z-boson production~\cite{Dubovyk:2016zok}.

For computations beyond the traditional use cases of \secdec{}, all individual parts of the algebraic manipulation are
separately available to the user. For example, it is possible to have \pysecdec{} only perform the Feynman parametrization
and further process it independently.

As part of \pysecdec{}, we provide the \cpp{} library \texttt{secdecutil}.
Class templates that are useful in a more general context than sector decomposition are implemented therein,
for instance a class template that defines addition, subtraction, multiplication, and division of series expansions.
Users are encouraged to use the low-level functionalities of \pysecdec{} to write their own high-level routines suitable
for the problem at hand.

%%%%%%%%%%%%%%%
\section{Usage}
%%%%%%%%%%%%%%%
\label{sec:usage}

The typical usage of \pysecdec{} is to compute integrals of the form
\begin{equation}\label{eq:generic_parameter_integral}
    I = p_0\left(\{a\},\{\epsilon\}\right)\int_0^1 \mathrm{d}x_1 ... \int_0^1 \mathrm{d}x_N \prod_{i=1}^M f_i\left( \{x\}, \{a\} \right) ^ {b_i + \sum_k c_{ik} \epsilon_k},
\end{equation}
where $p_0$ is an arbitrary prefactor, the $f_i$ are polynomials in $\{x\} = \lbrace x_1,...,x_N \rbrace$ with coefficients $\{a\}$,
the $b_i$ and $c_i$ are numeric constants, and the $\epsilon_k$ are analytic regulators.
Typically, $I$ has regulated divergences such that $\left. I \right| _ {\epsilon_k = 0} = \infty$.
A generic (multi-)loop integral of the form
\begin{eqnarray}\label{eq:loop_integral}
G &=& \int\prod\limits_{l=1}^{L} \mathrm{d}^D\kappa_l\;
\frac{1}
{\prod\limits_{j=1}^{N} P_{j}^{\nu_j}(\{k\},\{p\},m_j^2)}\nonumber\\
\mathrm{d}^D\kappa_l&=&\frac{\mu^{4-D}}{i\pi^{\frac{D}{2}}}\,\mathrm{d}^D k_l\;,\;
P_j(\{k\},\{p\},m_j^2)=(q_j^2-m_j^2+i\delta)\;,
\end{eqnarray}
where the $q_j$ are linear combinations of external momenta $p_i$ and loop momenta $k_l$,
can be brought to a parametric form as in equation~(\ref{eq:generic_parameter_integral}),
\begin{equation}\label{eq:param_rep}
G = (-1)^{N_{\nu}}
\frac{\Gamma(N_{\nu}-LD/2)}{\prod_{j=1}^{N}\Gamma(\nu_j)}\int
\limits_{0}^{\infty}
\,\prod\limits_{j=1}^{N}dx_j\,\,x_j^{\nu_j-1}\,\delta(1-\sum_{l=1}^N x_l)\,\frac{{\cal U}^{N_{\nu}-(L+1) D/2}}
{{\cal F}^{N_\nu-L D/2}} \;,
\end{equation}
where $N_\nu = \sum_j \nu_j$, by Feynman parametrization. This step is implemented in \pysecdec{}; i.e. integrals like in equation~(\ref{eq:loop_integral})
can directly be passed to \pysecdec{}. \pysecdec{} also implements Feynman parametrization of tensor integrals, which is omitted in the equations above for brevity.

The result is presented as series expansion in the analytic regulators
%, e.g.
%\begin{equation*}
%    \sum_{n=n_{min}}^{n_{max}} C_n \; \epsilon^n
%\end{equation*}
%for a single regulator $\epsilon$,
with numerical coefficients.
Obtaining a result with \pysecdec{} generally takes three steps: (i) Define the integral, (ii) generate and build the \cpp{} library for numerical integration,
(iii) insert numerical values for the parameters $\{a\}$ (Mandelstam invariants for loop integrals) and perform the numerical integration. In the following,
we guide through the most important matters on the way to a numerical result.

\subsection{Definition of an integral}
%%%%%%%%%%%%%%%%%%%%%%%%%%%%%%%%%%%%%%

\begin{figure}[htb]
    \vspace{5mm}
    \centering
    \begin{subfigure}[b]{0.45\textwidth}
        \includegraphics[width=\textwidth]{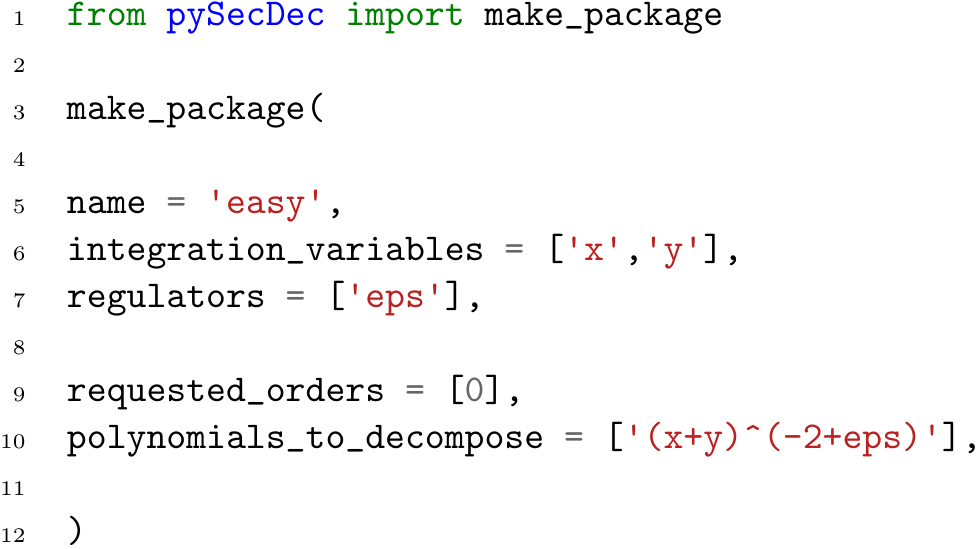}
        \vspace{1.8cm}
        \caption{}
        \label{code:minimal_generic}
    \end{subfigure}
    \qquad %add desired spacing between images, e. g. ~, \quad, \qquad, \hfill etc.
    \begin{subfigure}[b]{0.45\textwidth}
        \includegraphics[width=\textwidth]{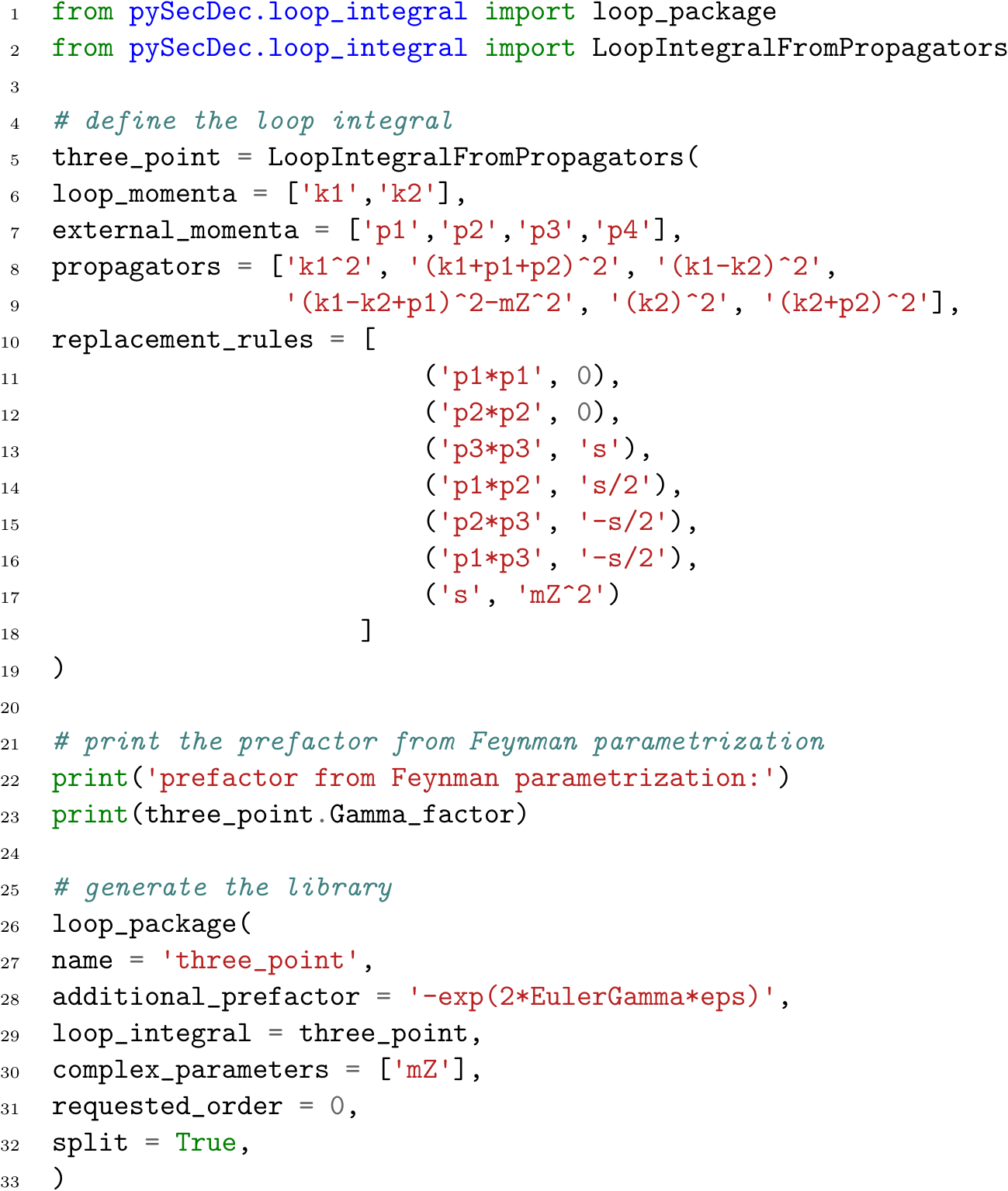}
        \caption{}
        \label{code:Zbb}
    \end{subfigure}
    \caption{Example runcards for \pysecdec{}. (a) Runcard to compute the integral defined in equation~(\ref{eq:easy}).
        (b) Runcard to compute the three-point function depicted in Figure~\ref{fig:Zbb}.}
    \label{code:minimal_runcards}
\end{figure}

Integrals are communicated to \pysecdec{} by running a python script, often referred to as "runcard",
that calls \texttt{make\_package} (\texttt{loop\_package} for loop integrals).
A minimal input to \texttt{make\_package} consists of an integral name, the integration variables,
the regulators, the maximum orders to expand in the regulators, and the integrand. Figure~\ref{code:Zbb}
shows a minimal runcard to compute
\begin{equation}\label{eq:easy}
    \int_0^1 \mathrm{d}x \; \int_0^1 \mathrm{d}y \;\, (x+y)^{-2+\epsilon}
\end{equation}
up to finite order in $\epsilon$. The integrand can be given as a list of the polynomials $f_i$ in the argument ``polynomials\_to\_decompose''.
If some of the input polynomials don't give rise to divergences, e.g. numerators of loop integrals, they can be
passed via the argument ``other\_polynomials'' instead. \pysecdec{} fully decomposes the ``polynomials\_to\_decompose'' but does
not force full decomposition of the ``other\_polynomials''. As a consequence, less sectors and therefore less integrand functions are generated.
It may, however, be beneficial to decompose such polynomials anyways.
The reason is that decomposed polynomials can have a lower intrinsic variance which is proportional to the error of any Monte Carlo
integrator. It strongly depends on the integral at hand if decomposing these polynomials yields better results.
If your integral has poles originating from the endpoint $1$, for example from polynomials like $(1-x)$ or $(x-y)$, the ``split'' option should be
set to ``True''.

A runcard to compute a loop integral consists of two steps. Fist, the loop integral is defined by either instantiating
\texttt{LoopIntegralFromPropagators} or \texttt{LoopIntegralFromGraph}. These classes perform the
Feynman parametrization given a list of propagators or a graph, respectively.
Advanced \python{} programmers can use these classes in their own code, for example to analytically process the Symanzik polynomials
$\U$ and $\F$. We refer the reader to the package documentation for a full description of Feynman parametrization in \pysecdec{}.
Second, the library for numerical integration is generated by a call to ``loop\_package''.
An example runcard how to print the Feynman prefactor $(-1)^{N_{\nu}}{\Gamma(N_{\nu}-LD/2)}/{\prod_{j=1}^{N}\Gamma(\nu_j)}$ and generate
the integration library of the three-point function depicted in Figure~\ref{fig:Zbb} is shown in Figure~\ref{code:Zbb}.
Note that we set ``split'' to ``True'' in our example which is required due to the special kinematics $s=m_Z^2$.
Further note that $m_Z$ is declared as complex parameter allowing for a nonzero width. We also define an ``additional\_prefactor''
which in our example means we compute $-\exp(2 \gamma_E \epsilon) G$ where $G$ is the three-point function. This is a change of convention
compared to \secdecthree{} where the additional prefactor was divided out rather than multiplied.
The resulting ``prefactor'' visible in the numerics (see next section) is the product of the Feynman prefactor and the
``additional\_prefactor'' provided by the user. By default, the ``additional\_prefactor'' is one and therefore the prefactor in the numerics
defaults to the Feynman prefactor.

\subsection{Numerical integration}
%%%%%%%%%%%%%%%%%%%%%%%%%%%%%%%%%%

The following description applies for both, libraries generated with \texttt{make\_package} and with \texttt{loop\_package}.
In fact, \texttt{loop\_package} internally calls \texttt{make\_package} after Feynman parametrization. Therefore, the
distinction between the loop and the general branch present in \secdecthree{} no longer exists and consequently all features that make sense
in both cases are available in both.

To generate the library for numerical integration, run \python{} on the runcard first, then run \texttt{make} in the generated directory.
For example, if you save the code of Figure~\ref{code:Zbb} to a file called \texttt{generate.py}, run \texttt{python generate.py}.
That generates a directory specified by the ``name'' argument (``easy'' in our example). To build the library run \texttt{make -C easy}.
The level of paralellization for building the library can be controlled from the command line via the environment variable \texttt{FORMTHREADS} (default: 2)
and the \texttt{-j} option of \texttt{make}. The variable \texttt{FORMTHREADS} is forwarded as \texttt{-w} option in every
invocation of \form{}; i.e. it controls how many threads \form{} uses to process each single sector. The \texttt{-j} option of bash controls how many instances
of \form{} and the compiler are launched in paralell; i.e. how many sectors are processessed at the same time. For example, to build the library using up to 4 processes,
and 3 \form{} threads per process run \texttt{make -j4 FORMTHREADS=3 -C easy}.

We wrap the \CUBA{}~\cite{Hahn:2004fe,Hahn:2014fua} library and \texttt{cquad} as implemented in the GNU scientific library (GSL)~\cite{Gough:2009:GSL:1538674}
for easy use with \pysecdec{}. Both packages are redistributed with \pysecdec{} and automatically available on systems where \pysecdec{} is properly installed.

\begin{figure}[htb]
\vspace{5mm}
\begin{center}
    \begin{minipage}{0.75\textwidth}
        \includegraphics[width=\textwidth]{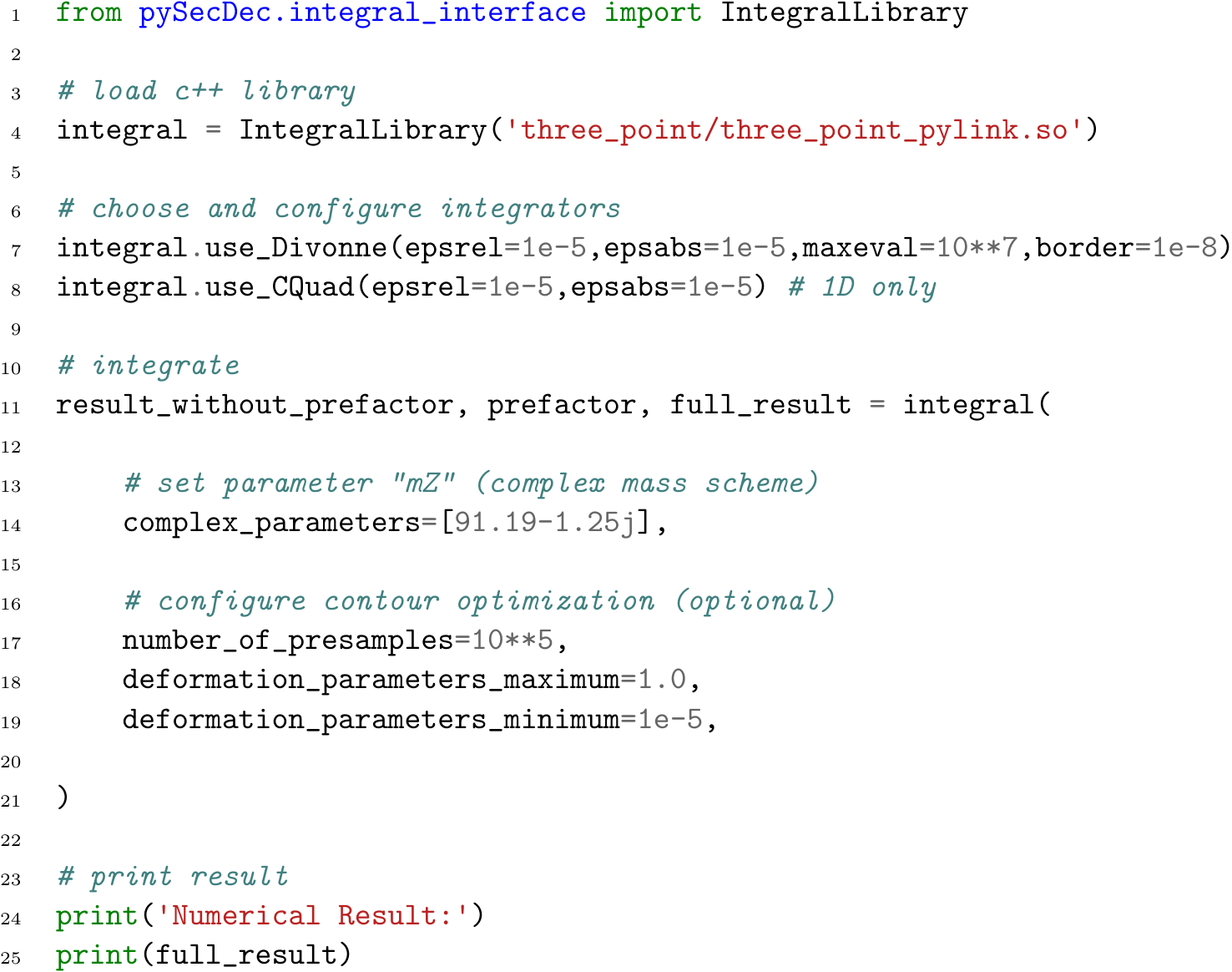}
    \end{minipage}
    \caption{Example script to issue the numerical integration using the python interface.}
    \label{code:Zbb_integrate}
\end{center}
\end{figure}

The easiest way to issue the numerical integration is the \python{} interface. The integration script shown in
Figure~\ref{code:Zbb_integrate} loads the necessary libraries, configures the integrators, runs the contour optimization and
numerical integration, and prints the result. To run the numerical integration of your integral, write an integration script analogous to
the one depicted in Figure~\ref{code:Zbb_integrate}. Assuming the integration script is called \texttt{integrate.py}, running
\texttt{python integrate.py} will perform the integration and print the result to the screen.

The numerical accuracy goals of all integrators are controlled by the options ``epsrel'' and ``epsabs''. In addition, \CUBA{} integrators
also implement ``maxeval''. The integration terminates when either goal is reached. If your integral evaluates to a very small value, make
sure that you decrease ``epsabs'' accordingly. The integrator \texttt{CQuad} is restricted to one dimensional integrands.
A call to \texttt{integral.use\_CQuad} makes \pysecdec{} call \texttt{CQuad} for every one dimensional integral
while the integrator that was assigned before is used for higher dimensional integrals. Without call to \texttt{integral.use\_CQuad},
the \CUBA{} integrator is also used for one dimensional integrals. When using a \CUBA{} integrator, the number of threads
used for the integration is controlled by the environment variable \texttt{CUBACORES}. For example, to run with 6 threads invoke
the script as \texttt{CUBACORES=6 python integrate\_triangle2L.py}.

The parameters left symbolic in the generation step must be set to numerical values when invoking the numerical integration.
The call to \texttt{integral} takes a list of ``real\_paramters'' and ``complex\_parameters''. If there are no real or complex
parameters, the argument may be omitted (``real\_paramters'' in our example). Complex numbers (the Z-boson mass \texttt{91.19-1.25j}
in our example) are entered as \#$\pm$\#j in the python interface. The call that invokes the main integration also takes parameters that
guide the optimization of the deformed contour. These parameters should usually be omitted; i.e. left at their default values unless the
integration terminates due to a ``sign\_check\_error''.

A ``sign\_check\_error'' means that the chosen contour in the analytic continuation to the complex plane
is invalid. Since the contour deformation is by construction~\cite{Soper:1999xk,Binoth:2005ff,Anastasiou:2007qb,Borowka:2012yc,Schlenk:2016cwf} guaranteed to be valid
with sufficiently small deformation parameters, you should try to decrease ``deformation\_parameters\_maximum''. We suggest to repeatedly lower ``de\-for\-ma\-tion\_pa\-ra\-me\-ters\_max\-i\-mum''
by an factor of ten until the integration succeeds. If your integral evaluates to a very small number, you may have to decrease the ``deformation\_parameters\_maximum''
below 1e-5 which is the default value of ``deformation\_parameters\_minimum''. If that happens, make sure that you also decrease ``deformation\_parameters\_minimum''
to keep it smaller than the maximum.

Note that often large cancellations giving rise to numerical instabilities occur in the integrand functions when the integration variables are close to zero. This can be
problematic for adaptive integrators, in particular Divonne. It is usually necessary to set a nonzero border when using Divonne to avoid samples in unstable boundary regions.
If you keep getting NaN, you should switch to Vegas and increase the default values for ``nincrease'' (default: $500$) and ``nstart'' (default: $1000$) by a factor of ten or
more. Starting from \pysecdec{} version 1.3, you can also use the new \texttt{zero\_border} option to avoid evaluation of the integrand close to the zero boundary.

%%%%%%%%%%%%%%%%%%%%
\section{Conclusion}
%%%%%%%%%%%%%%%%%%%%
\label{sec:conclusion}

We have presented \pysecdec{}, the successor of the \secdecthree{}, a program that numerically computes parameter integrals, e.g. (multi-)loop integrals after Feynman parametrization,
in the context of dimensional regularization. The open-source package can be downloaded from \href{http://secdec.hepforge.org/}{http://secdec.hepforge.org/}.

A key new feature is the output of a \cpp{} library, which is also accessible from \python{}. Existing sector decomposition programs are mainly designed to provide cross-checks against
analytic calculations. The different design goal makes dynamically embedding them into user-defined calculations a difficult task. However, there is foreseeable need for numerical solutions
of integrals for which analytical solutions will not become available in the near future. The situation motivated the development of a program that makes both use cases easy, usage as static
cross-check and as dynamic library.

\pysecdec{} is a tool that satisfies the aforementioned demands. It can dynamically provide numerical solutions of loop-integrals in user-defined codes, for example amplitude
calculations. Static cross-checks can be performed with a program that evaluates the integral at a single point. It turns out that writing such a program is as easy as writing a runcard
with a suitable library interface. Nevertheless, a library opens the door to a much wider field of use cases compared to runcards.

We have demonstrated how integrals are communicated to \pysecdec{} and explained how to write \python{} scripts that evaluate the integral using the \python{} interface. Note that the
scripts shown in this article can be embedded into arbitrary user-defined \python{} code. However, the \python{} interface provides only a fraction of the library's functionality.
We therefore recommend to switch to the more complex but much more powerful \cpp{} interface for dynamical library use.

%%%%%%%%%%%%%%%%%%%%%%%%%%
\section*{Acknowledgments}
%%%%%%%%%%%%%%%%%%%%%%%%%%

I thank my fellow \secdec{} developer team, Sophia Borowka, Gudrun Heinrich, Stephen Jones, Matthias Kerner, Johannes Schlenk,
and Tom Zirke, for helpful discussions and the successful collaboration.

\bibliographystyle{JHEP}
\bibliography{refs_secdec}

\end{document}